\documentclass[aps,prd,groupedaddress,nofootinbib]{revtex4}

\def\ga{\mathrel{\raise.3ex\hbox{$>$\kern-.75em\lower1ex\hbox{$\sim$}}}}
\def\la{\mathrel{\raise.3ex\hbox{$<$\kern-.75em\lower1ex\hbox{$\sim$}}}}

\def\lsim{\mathrel{\rlap{\lower4pt\hbox{\hskip1pt$\sim$}}
    \raise1pt\hbox{$<$}}}                
\def\gsim{\mathrel{\rlap{\lower4pt\hbox{\hskip1pt$\sim$}}
    \raise1pt\hbox{$>$}}}                

\usepackage{amsmath}
\usepackage{amssymb}
\usepackage{bbm}
\usepackage{epsfig}
\usepackage{slashed}

\newcommand{\ba}{\begin{array}}
\newcommand{\ea}{\end{array}}
\newcommand{\beq}{\begin{equation}}
\newcommand{\eeq}{\end{equation}}
\newcommand{\comment}[1]{}
\def\bt{\begin{table}}
\def\et{\end{table}}
\def\bc{\begin{center}}
\def\ec{\end{center}}
\def\bi{\begin{itemize}}
\def\ei{\end{itemize}}
\def\bea{\begin{eqnarray}}
\def\eea{\end{eqnarray}}
\def\beas{\begin{eqnarray*}}
\def\eeas{\end{eqnarray*}}

\begin{document}

\title{\boldmath New physics contributions to $A^{t \bar t}_{FB}$ at the Tevatron
}
\author{Sudhansu S. Biswal$^{1}$~\footnote{sudhansu.biswal@gmail.com}, Subhadip Mitra$^{2}$~\footnote{smitra@iopb.res.in},  Rui Santos$^{3,4}$~\footnote{rsantos@cii.fc.ul.pt},  Pankaj Sharma$^{5}$~\footnote{pankajs@prl.res.in}, Ritesh K. Singh$^{6}$~\footnote{ritesh.singh@iiserkol.ac.in} and Miguel Won$^{7}$~\footnote{miguel.won@coimbra.lip.pt}}
\affiliation{$^1$Department of Physics, Orissa University of Agriculture and Technology, Bhubaneswar 751003, India.}
\affiliation{$^2$Institute of Physics, Bhubaneswar 751005, India.}
\affiliation{$^3$Instituto Superior de Engenharia de Lisboa, Rua Conselheiro Em\'\i dio Navarro 1, 1959-007 Lisboa, Portugal.}
\affiliation{$^4$Centro de F\'\i sica Te\' orica e Computacional, Faculdade de Ci\^encias, Universidade de Lisboa, Av. Prof. Gama Pinto 2, 1649-003 Lisboa, Portugal.}
\affiliation{$^5$Physical Research Laboratory, Ahmedabad, India.}
\affiliation{$^6$Department of Physical Sciences, 
Indian Institute of Science Education and Research -- Kolkata,
Mohanpur campus, 741252, West Bengal, India}
\affiliation{$^7$LIP - Departamento de F\'isica, Universidade de Coimbra, Coimbra, Portugal.}

\date{\today}

\begin{abstract}
The Tevatron has measured a discrepancy relative to the Standard Model 
prediction  in the forward-backward asymmetry in top quark pair production. 
This asymmetry grows with the rapidity difference of the two top quarks. It 
also increases with the invariant mass of the tt pair, reaching, for high 
invariant masses, 3.4 standard deviations above the Next to Leading Order prediction for the 
charge asymmetry of QCD. However, perfect agreement between experiment and the
Standard Model was found in both total and differential cross section of top
quark pair production. As this result could be a sign of new physics we have 
parametrized this new physics in terms of a complete set of dimension six 
operators involving the top quark. We have then used  a Markov Chain Monte 
Carlo approach in order to find the best set of parameters that fits the data,
using all available data regarding top quark pair production at the Tevatron.
We have found that just a very small number of operators are able to fit the
data better than the Standard Model.
\end{abstract}


\maketitle

\section{Introduction}


\noindent
The most recent measurement of the forward-backward asymmetry, $A^{t\bar t}_{FB}$,  in top 
quark pair production at the Tevatron~\cite{:2007qb, Aaltonen:2008hc} was 
performed by the CDF collaboration using a data sample with 5.3 fb$^{-1}$ of 
integrated luminosity~\cite{Aaltonen:2011kc}. After background subtraction, 
the value of $A^{t\bar t}_{FB}$ in the center-of-mass (CM) frame of the top quarks is
\begin{equation}
A^{t \bar t}_{FB} = 0.158 \pm 0.074
\end{equation}
which constitutes about two standard deviations above the Next-to-Leading-Order
(NLO) Standard Model (SM) prediction~\cite{Kuhn:1998jr}
\begin{equation}
A^{t \bar t, \rm{SM}}_{FB}    = 0.058 \pm 0.009 \quad .
\end{equation}
Despite the discrepancy in $A^{t\bar t}_{FB}$, the total $t 
\bar t$ production cross section  is in good agreement with the SM prediction. In fact, with 
4.6 fb$^{-1}$ collected luminosity, the top quark pair production cross 
section~\cite{Aaltonen:2010ic} yields the result
\begin{equation}
\sigma^{\rm{Measured}}_{t \bar{t}} =7.70 \pm 0.52 \quad pb
\end{equation}
for a top quark of mass 172.5 GeV, which is in good agreement with the 
theoretical prediction~\cite{Campbell:1999ah}
\begin{equation}
\sigma^{\rm{SM}}_{t \bar{t}}(\rm{MCFM}) =7.45^{+0.72}_{-0.63} \quad pb
\end{equation}
where MCFM stands for Monte Carlo for FeMtobarn processes~\cite{site}.
Measurements of the $t \bar t$ differential cross section with the $t \bar t$ 
invariant  mass ($m_{t\bar t}$), $d \sigma / d m_{t \bar t}$ were also performed by the CDF 
collaboration~\cite{cxdifCDF}. With an integrated luminosity of 2.7 fb$^{-1}$ 
CDF has tested the $m_{t \bar t}$ spectrum for consistency with the SM. The 
results are presented in table \ref{tab:invmass}. They have concluded that 
there is no evidence of non-SM physics in $m_{t \bar t}$ distributions. 
Hence, whatever new physics explains the forward-backward asymmetry in $t \bar t$ 
production, it has to comply with all other measurements that are in agreement 
with the SM. Finally, measurements of the asymmetry for two regions of the 
top-antitop rapidity difference ($\Delta Y$) and for two regions of the 
invariant mass ($m_{t \bar{t}}$) were performed by the CDF collaboration 
in~\cite{Aaltonen:2011kc}. The results are presented in table~\ref{tab:Ymass} 
together with the theoretical predictions .
The asymmetry at high mass is 3.4 standard deviations above the NLO prediction 
for the charge asymmetry of QCD. Recently the electroweak contributions to the 
asymmetry were re-analysed~\cite{Hollik:2011ps, Kuhn:2011ri} just to conclude 
that the observed mass-dependent forward-backward asymmetry still shows a 
3$\sigma$ deviation in the high mass region. 
Inclusion of corrections beyond NLO does not  change this picture as well~\cite{Ahrens:2011uf}.
The separate results at high mass 
and large $\Delta Y$ contain partially independent information on the asymmetry
mechanism. Therefore, a total of 14 observables were measured at the Tevatron. 
This set of experimental values will be used to investigate whether the 
complete set of effective dimension six operators is able to describe the 
possible new physics responsible for the observed discrepancies while retaining
the measurements in agreement with the SM. Recently D0~\cite{Abazov:2011rq} has
measured $A^{t\bar t}_{FB}$ with 5.4 fb$^{-1}$ of collected 
luminosity. As discussed in~\cite{Grinstein:2011dz}, their analysis
does not observe a significant rise of the “folded” detector level asymmetry
with respect to $m_{t \bar{t}}-$ and $\Delta Y$. Until
these results are unfolded they can not be directly compared to the CDF
ones, even if at the detector level they
appear to be consistent within errors.
We did not use the results~\cite{Abazov:2011rq} in our analysis.

\begin{table}[ht!]
\begin{center}
\begin{tabular}{c c c}
\hline
\hline
Bin     &   $\sigma$ (CDF result)  & $\sigma$ (SM-NLO) \\
(GeV)   &   (pbarn) &  (pbarn)\\
\hline
350-400          &  $3.115  \pm 0.559$ &  $2.450$       \\
400-450          &  $1.690 \pm 0.269$ &  $1.900$       \\
450-500          &  $0.790 \pm 0.170$ &  $1.150$       \\
500-550          &  $0.495 \pm 0.114$ &  $0.600$       \\
550-600          &  $0.285 \pm 0.071$ &  $0.400$       \\
600-700          &  $0.230 \pm 0.073$ &  $0.310$       \\
700-800          &  $0.080 \pm 0.037$ &  $0.100$       \\
800-1400         &  $0.041 \pm 0.021$ &  $0.036$       \\
\hline
\hline
\end{tabular}
\caption{CDF measurements of $d\sigma/dm_{t \bar{t}}$~\cite{cxdifCDF} (integrated in each bin). We 
bin-wise scale our SM result (at LO) to match the SM-NLO result to emulate a 
$m_{t\bar t}$ dependent $k$-factor for fitting. The SM-NLO values are extracted
from the plot in~\cite{Cao:2010zb}.
\label{tab:invmass}}
\end{center}
\end{table}

\begin{table}[ht!]
\begin{center}
\begin{tabular}{lrr}\hline\hline
Observables & CDF result \; \; & SM prediction \; \; \\ \hline
$A_{FB}^{t \bar{t}}(|\Delta Y_t|<1.0)$ & $(0.026\pm 0.118)$ &
$(0.039\pm 0.006)$ \\
$A_{FB}^{t \bar{t}}(|\Delta Y_t|>1.0)$ & $(0.611\pm 0.256)$ &
$(0.123\pm 0.008)$ \\
$A^t_{FB}(m_{t\bar t}-)$ & $(-0.116\pm 0.153)$ &
$(0.040\pm 0.006)$\\
$A^t_{FB}(m_{t\bar t}+)$ & $(0.475\pm 0.114)$ &
$(0.088\pm 0.013)$\\ \hline
\end{tabular}
\end{center}
\caption{CDF measurements~\cite{Aaltonen:2011kc} and SM predictions for the 
Forward-Backward Asymmetry for two regions of $\Delta Y_t$ and for two regions
of $m_{t\bar{t}}$. $A^t_{FB}(m_{t\bar t}+)$ stands for $A_{FB}^{t \bar{t}}(m_{t\bar{t}} > 450~{\rm GeV})$,
while $A^t_{FB}(m_{t\bar t}-)$ stands for $A_{FB}^{t \bar{t}}(m_{t\bar{t}} < 450~{\rm GeV})$. \label{tab:Ymass}}
\end{table}

\noindent
There have been several attempts to explain this discrepancy. The most popular 
collection of models among theorists when trying to account for the Tevatron 
results are the ones with new gauge bosons, and in particular, axigluons, 
W$'$ and Z$'$ bosons~\cite{Gabrielli:2011zw}. 
Explanations in the framework of SuperSymmetric models were discussed 
in~\cite{Isidori:2011dp}. Other possible justifications for the 
inconsistency between theory and experiment in the asymmetry, while leaving 
the cross section for $t \bar{t}$ production within measured uncertainties, 
include $t$-channel exchange of color sextet or triplet scalar 
particles~\cite{Shu:2009xf}, $s$-channel coloured unparticle 
contributions~\cite{Chen:2010hm} or $s$-channel new colour octet vector bosons
contributions~\cite{Wang:2011hc}, light coloured particles  from a particular SU(5) GUT 
model~\cite{Dorsner:2009mq}, extra dimensions~\cite{Djouadi:2011aj}, 
SO(10) models~\cite{Patel:2011eh}, SO(5)~$\otimes$~U(1) 
gauge-Higgs unification models~\cite{Uekusa:2009qx}, new heavy 
quarks~\cite{Davoudiasl:2011tv}, diquark models~\cite{Arhrib:2009hu} and models
where $SU(3)_c$ QCD theory is extended to $SU(N)_c$ which is spontaneously 
broken at a scale just above the weak scale~\cite{Foot:2011xu}.

\noindent
The search for resonances decaying into $t \bar t$ has also been carried out at 
the Tevatron~\cite{Aaltonen:2011ts} (see also~\cite{Aaltonen:2011vi}) with 
negative results. CDF has tested vector resonances with masses between 450 GeV 
and 1500 GeV with widths equal to 1.2 \% of their mass. With 4.8  fb$^{-1}$ of 
integrated luminosity they found no evidence of resonant production of $t \bar 
t $ candidate events. This result sustains the argument of integrating out new 
heavy fields and strengthens the idea of adopting a model independent approach 
in explaining the measured asymmetry at the Tevatron. An independent approach, 
with the recourse to higher dimension operators was already discussed 
in~\cite{AguilarSaavedra:2011ci}. In this 
work we propose to study the effect of dimension six flavour changing neutral 
current (FCNC) operators together with four fermion (4F) operators. In order to find
the best set of parameters that fits the data we will use a Markov Chain Monte 
Carlo (MCMC) approach.  

\noindent
The paper is organized as follows. The next section is devoted to describe the 
effective operator approach and the number of independent operators that will 
be used in the analysis. In section 3 we describe the MCMC method and we 
present the results for the complete set of operators. Finally a discussion on 
the results and the conclusions are presented in section 4.

\section{The effective operator approach}
\label{sec:eff}

\noindent
The Standard Model of particle physics is the low energy limit of a more 
general theory which could manifest itself through a set of effective operators
of dimensions higher than four. 
The effective operator formalism assumes that this more general theory would be 
visible at very high energies and at an energy scale lower than $\Lambda$, the 
set of higher order operators would be suppressed by powers of $\Lambda$.
The Lagrangian of the new theory can be written as a series in  $\Lambda$ with 
operators obeying  the gauge symmetries of the SM
\begin{equation}
{\cal L} \;\;=\;\; {\cal L}^{SM} \;+\; \frac{1}{\Lambda}\,{\cal
L}^{(5)} \;+\; \frac{1}{\Lambda^2}\,{\cal L}^{(6)} \;+\;
O\,\left(\frac{1}{\Lambda^3}\right) \;\;\; , \label{eq:l}
\end{equation}
where ${\cal L}^{SM}$ is the SM lagrangian and ${\cal L}^{(5)}$ and 
${\cal L}^{(6)}$ contain all the dimension five and six operators respectively.
This formalism allow us to parametrize new physics, beyond that of the SM, in 
a model-independent manner. The term $\mathcal L^{(5)}$ is eliminated by baryon 
and lepton number conservation. Thus, any new particle or interaction is hidden 
in the dimension six operators which are listed in~\cite{buch, 
Grzadkowski:2010es}.

\noindent
We divide the dimension six operators in two groups, the four-fermion (4F) 
operators and the non-4F operators. The later can then be grouped according to 
the gauge boson present in the triple vertex. As we are discussing $t \bar t$ 
production, the non-4F operators contributing to the process have at least one 
top quark in the interaction. Operators with one top quark, a light up-quark 
and one gauge boson will be called FCNC 
operators. If the gauge boson is a gluon they are classified as strong FCNC 
operators~\cite{Ferreira:2005dr, Ferreira:2006xe}; otherwise they will be 
called electroweak FCNC operators~\cite{Ferreira:2008cj, Coimbra:2008qp}.

\begin{figure}[h!]
\epsfysize=3.cm \centerline{\epsfbox{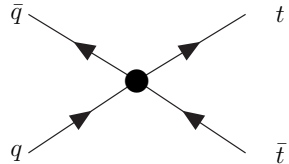}}
\caption{Four-fermion Feynman diagrams for $t \, \bar{t}$ production.}
\label{fig:4F}
\end{figure}

\noindent
When looking for new physics that would explain the $t \bar t$ asymmetry in 
the framework of the effective operator approach we start by looking at 
the dimension six non-FCNC operators. As the final state is $t \bar t$, the only 
possible new vertex is an anomalous $g t \bar t$ interaction. Its contribution 
to the process would originate from the diagrams
\begin{figure}[h!]
\epsfysize=3.cm \centerline{\epsfbox{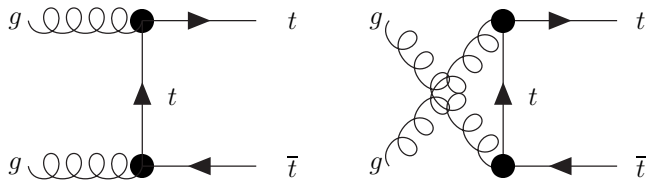}}
\caption{Feynman diagrams for $t \, \bar{t}$ production via anomalous $g t \bar t$ interaction.}
\label{fig:ggtt}
\end{figure}
presented in Fig.~\ref{fig:ggtt}. However, because the initial state is 
symmetric, these diagrams will only contribute to the cross section but not to 
the asymmetry.  Therefore, any changes produced by these operators would change 
only the cross section, but not the asymmetry where the discrepancy is.
The next class of operators we discuss are the FCNC ones. In this case there 
are two sets of diagrams to consider: the ones initiated by gluons 
\begin{figure}[h!]
\epsfysize=3.cm \centerline{\epsfbox{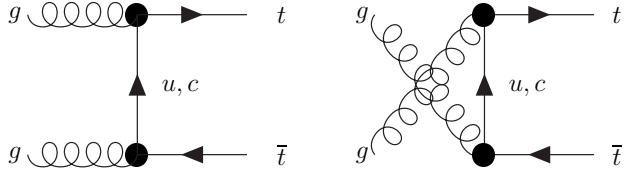}}
\caption{Feynman diagrams with FCNC operators for $t \, \bar{t}$ production 
via gluon fusion.}
\label{fig:ggttFCNC}
\end{figure}
and the ones initiated by light quarks.
\begin{figure}[h!]
\epsfysize=3.cm \centerline{\epsfbox{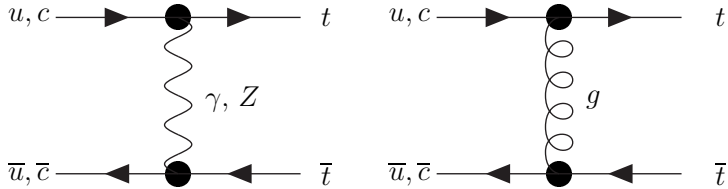}}
\caption{FCNC Feynman diagrams for $t \, \bar{t}$ production via $q \bar q$ fusion.}
\label{fig:qqttFCNC}
\end{figure}
For the same reasons discussed for Fig.~\ref{fig:ggtt} the contribution of 
the diagrams in Fig.~\ref{fig:ggttFCNC} to the asymmetry is negligible. 
Therefore there are only contributions coming from the diagrams in 
Fig.~\ref{fig:qqttFCNC}. Regarding those diagrams 
(Fig.~\ref{fig:qqttFCNC}), and taking into account that the contribution of 
the c-quark is much smaller than that of the u-quark, we discard all 
contributions that have a c quark in the initial state. Note that there are no 
s-channel contributions for the FCNC case because we have a top-antitop pair 
in the final state. Finally we will consider all 4F-fermion operators as shown 
in Fig.~\ref{fig:4F}. 
Note that we will consider not only the interference term with the SM 
contribution, of order $1/\Lambda^2$ but also the modulus-square terms of order 
$1/\Lambda^4$. We will now discuss the minimum number of operators to be used 
in our analysis.    

\subsection{Effective operators in the strong sector}
\noindent
Following the notation of~\cite{buch}, the operators contributing to the strong 
FCNC vertices can be written as
\begin{equation}
{\cal O}^{ij}_{uG \phi} \, = \,  \bar{q}^{i}_L \, \lambda^{a} \, \sigma^{\mu\nu} \, u^{j}_R \, \tilde{\phi} \, G^{a \mu \nu}  \,  ,
\label{eq:op1}
\end{equation}
and
\begin{equation}
{\cal O}^{ij}_{uG} \, = \,  \bar{u}^i_R \, \lambda^{a} \, \gamma_{\mu}  D_{\nu} u^j_R \, G^{a \mu \nu}  \, , \quad 
{\cal O}^{ij}_{qG} \, = \,   \bar{q}^i_L \,
\lambda^{a} \, \gamma_{\mu}  D_{\nu} q^j_L \, G^{a \mu \nu} \, ,
\label{eq:op2}
\end{equation}
where $G^a_{\mu\nu}$ is the gluonic field tensor, $u^i_R$ stands for a 
right-handed quark singlet and $q^i_L$ represents the left-handed quark 
doublet. FCNC occurs because one of the indices is always equal to 3 while the 
other is either 1 or 2, that is, there is always one (and one only) top-quark 
present in the operator; the remaining fermion field in the interaction is 
either a u or a c-quark. Throughout this section we assume that 
$\mathcal O^{ij}$ and $\mathcal O^{ji}$ are independent operators and the 
hermitian conjugate of all the operators are included in the final Lagrangian.
The operators in~(\ref{eq:op1}) are related to the operators in~(\ref{eq:op2}) 
through equations of motion that also involve 4F operators~\cite{buch, 
Grzadkowski:2010es, Ferreira:2005dr, Ferreira:2006xe, AguilarSaavedra:2008zc}. 
However, the 4F operators appearing in those equations have either one or three 
top-quarks~\cite{AguilarSaavedra:2008zc}. Therefore, if those  4F operators can 
be discarded, operators  in~(\ref{eq:op2}) can be discarded as well. The 
operators presented in this section will give rise to the FCNC vertices of the 
form $g\,t\,\bar{u_i}$ (with $u_i \,=\,u\,,\,c$) and the corresponding 
hermitian conjugate interaction with an independent coefficient. 
\subsection{Effective operators in the electroweak sector}
\noindent 
There are also effective operators stemming from the electroweak sector that 
would give rise to new FCNC interactions involving the 
top quark~\cite{Ferreira:2008cj,Coimbra:2008qp}.  We start by listing the 
chirality flipping operators which are the equivalent to the ones in the strong 
sector, the only difference being the gluonic tensor replaced by the U(1) and 
SU(2) field tensors. They can be written as
\begin{equation}
{\cal O}^{ij}_{uB \phi}  \, = \, \bar{q}^i_L \, \sigma^{\mu\nu} \, u^j_R \, \tilde{\phi} \,B_{\mu\nu} \, , \quad
{\cal O}^{ij}_{uW \phi} \, = \, \bar{q}^i_L \, \tau_{I} \, \sigma^{\mu\nu} \, u^j_R \, \tilde{\phi} \, W^{I}_{\mu\nu} \,  ,
\label{eq:op3}
\end{equation}
and
\begin{equation}
{\cal O}^{ij}_{uB} \, = \, \bar{u}^i_R \, \gamma_{\mu}  D_{\nu} u^j_R \, B^{\mu \nu}  \, , \quad 
{\cal O}^{ij}_{qB} \, = \,  \bar{q}^i_L \, \gamma_{\mu}  D_{\nu} q^j_L \, B^{\mu \nu}  \, , \quad
{\cal O}^{ij}_{uW} \, = \, \bar{q}^i_L \, \tau_{I}  \, \gamma_{\mu}   D_{\nu} \, q^j_L \, W^{I}_{\mu\nu} \, ,
\label{eq:op4}
\end{equation}
%
where $B^{\mu \nu}$ and $W^{I}_{\mu\nu}$ are the $U(1)_Y$ and $SU(2)_L$ field 
tensors, respectively. There are also equations of motion in the electroweak 
sector that relate the operators in~(\ref{eq:op3}) with the ones  
in~(\ref{eq:op4}) and with 4F operators~\cite{AguilarSaavedra:2008zc}. A 
similar analysis to the one performed for the strong sector regarding the 
contribution of the 4F operators leads to the conclusion that we can neglect 
the operators (\ref{eq:op4}) in our analysis. 

\noindent
Besides chirality-flipping operators there are chirality conserving FCNC 
operators. Their flavour conserving counterparts are already present in the SM 
lagrangian at tree-level. In fact, the vertex $\bar{t} t Z$ has two vector 
contributions of different magnitudes, one proportional to $\gamma_{\mu}\, 
\gamma_L$ and the other proportional to $\gamma_{\mu}\, \gamma_R$. Hence the 
flavour conserving contribution would modify the Z boson neutral current. All 
the chirality conserving operators involve the Higgs doublet. As the Higgs 
field is electrically neutral, there are more effective operators which will 
only contribute to new $Z$ FCNC interactions. This set of operators can be 
written as
\begin{equation}
{\cal O}^{ij}_{\phi u}  \, = \, i  \, (\phi^{\dagger}  D_{\mu} \phi) \, (\bar{u}^i_R \,  \gamma^{\mu}  \,  u^j_R) \, , 
\label{eq:op5}
\end{equation}
\begin{equation}
{\cal O}^{(1),ij}_{\phi q} \, = \, i \,  (\phi^{\dagger} D_{\mu} \phi) \, (\bar{q}^i_L   \, \gamma^{\mu}  \,  q^j_L)  \, , \quad
{\cal O}^{(3),ij}_{\phi q} \, = \, i \, (\phi^{\dagger} \, \tau_{I}  \, D_{\mu} \phi)  \,   (\bar{q}^i_L \, \gamma^{\mu} \, \tau_{I}  \,  q^j_L)  \, ,
\label{eq:op6}
\end{equation}
and 
\begin{equation}
{\cal O}^{ij}_{D_u}  \, = \,   (\bar{q}^i_L\, D^{\mu}\, u^j_R)  \,   D_{\mu} \tilde{\phi}  \, , \quad
{\cal O}^{ij}_{\bar{D}_u}  \, = \, (D^{\mu} \bar{q}^i_L\, u^j_R ) \, D_{\mu} \tilde{\phi}    \, .
\label{eq:op7}
\end{equation}
%
Again, the use of the equations of motion allow us to discard the operators 
in~(\ref{eq:op7}). In the electroweak sector, there are now 4F operators with 
one top and one anti-top. However, those 4F operators always have one b-quark 
in the interaction or, if not, are CKM suppressed making its contribution to 
the $t \bar t$ asymmetry negligible. Furthermore, as was shown 
in~\cite{AguilarSaavedra:2008zc}, for all the operators in~(\ref{eq:op5}) and 
(\ref{eq:op6}),  ${\cal O}^{ij}$ and ${\cal O}^{ji}$ are not independent. This 
means that the number of independent operators in~(\ref{eq:op5}) and 
(\ref{eq:op6}) is reduced to three (for each light flavour). Finally, for this 
particular study, we can group  ${\cal O}^{(1),ij}_{\phi q}$ and 
${\cal O}^{(3),ij}_{\phi q} $ under the same Lorentz structure which further 
reduces the number of independent operators in~(\ref{eq:op6}) to two for each 
light flavour.

\noindent
The above discussion leads us to the conclusion that the minimum number of 
operators needed to describe the asymmetry is 8 for each light flavour. 

The amplitudes were generated with CalcHEP~\cite{Pukhov:1999gg}, 
the Feynman rules for the effective operators were derived with 
LanHEP~\cite{Semenov:1998eb} and the integration
was performed using the CUBA library~\cite{Hahn:2004fe}.

\subsection{Four-fermion operators}
\label{sec:4F}

\noindent
We now turn to the four-fermion operators. In order to make the analysis as 
clear as possible we will reduce the operators to a manageable number making 
use of all allowed reduction procedures, from equations of motion to Fierz 
identities. Again, because the largest contribution to $t\bar t$ production 
occurs in $u\bar u$ fusion, we will discard all non u-quarks contribution in 
our study. We end up with a total of 12 operators in agreement 
with~\cite{AguilarSaavedra:2010zi}, that is, 12 operators for each 
light up-quark flavour and we do not consider operators with down-quarks in 
the initial state. 
This simplification allow us to find hints of the type of operators that
can contribute to the asymmetry according to the its Lorentz structure.
We write the four fermion lagrangian as 
%
%
\begin{eqnarray}
{\cal L}_6^{4F} & = & \frac{g_s^2}{\Lambda^2}\sum_{A,B} \left[C_{AB}^{1} (\bar u_A
\gamma_\mu u_A)(\bar t_B\gamma^\mu t_B)
+C_{AB}^{8} (\bar u_AT^a\gamma_\mu u_A)(\bar t_BT^a\gamma^\mu t_B)  \right] + \nonumber\\
 && \frac{g_s^2}{\Lambda^2} \sum_{A \neq B} \left[ N_{AB}^{1} \,  (\bar u_A \gamma_\mu t_A)(\bar t_B\gamma^\mu u_B) 
 + N_{AB}^{8} \,  (\bar u_A T^a \gamma_\mu t_A)(\bar t_B T^a \gamma^\mu u_B) \right] \nonumber 
\label{eq:L6_4F}
\end{eqnarray}
where $T^a = \lambda^a/2$,  $\{A,B\} = \{L,R\}$, and the exponent $1$ and $8$ 
denotes a color singlet and a color octet interaction, respectively.

\section{Results}
\subsection{Parameter sampling method}
\noindent
In order to find the best set of parameters that fits the data we use a Markov 
Chain Monte Carlo (MCMC) approach. We start with a random point from the 
multi-dimensional parameter space for the chosen model. The $\chi^2$ for this
point is calculated and a likelihood is assigned to it. This likelihood is a measure 
of how well a set of data is reproduced for a given point in the model 
parameter space. Following the notation of~\cite{Belanger:2009ti}, this 
function is defined as
\begin{equation}
\textbf{G} (O, O_{exp}, \Delta O)=exp \left[  \frac{-\chi ^2 (O, O_{exp}, \Delta O)}{2} \right]
\end{equation}
where
\begin{equation}
\chi = \frac{O-O_{exp}}{\Delta O} 
\end{equation}
and $O$ is the value of the observable for a given point of parameter 
space, $O_{exp}$ is the central value of the observable and 
$\Delta O$ is the 1$\sigma$ error. The absolute value of the 
likelihood function is irrelevant for our analysis.  What is relevant is 
the ratio of likelihoods, that is, the comparison of the likelihoods of two 
consecutive points in the chain. We follow the Metropolis-Hastings (MH) 
algorithm - the Markov chain is started from a random initial value in 
parameter space with a given likelihood that depends on the constraints imposed 
by the data set. Next, a new point is generated randomly with a probability 
distribution centred around the old point and the likelihoods of the two 
points are compared - if the likelihood of the next point is larger than the 
one for the current point the next point is appended to the chain, otherwise 
the current point is replicated in the chain. 
We have repeated this procedure with 10 different random starting points. 
We scan with flat priors (i.e. a linear sampling over the parameters) and we 
have checked that the chains have a very good convergence behaviour. 
In all calculations of 
the top production cross sections we use the top mass as the renormalization 
and factorization scale. We take $m_t = 175$ GeV and to take into account the 
NLO corrections we have chosen a $k-$factor of 1.41~\cite{Campbell:1999ah,site}.
Further, we use a bin-wise scaling for the $m_{t\bar t}$ distribution to 
emulate the $m_{t\bar t}$ dependent $k-$factor.

\subsection{Strong and Electroweak operators}

\noindent
In order to simplify the notation we have replaced the original constants from 
the operators in the strong and electroweak sectors by $\alpha_i$ with $i=1,8$. 
This correspondence between constants $\alpha_i$ and the operators themselves 
is presented in table~\ref{tab:alphas}. As an example, the first operator 
$\mathcal O^{ut}_{uG\phi}$ would appear in the effective Lagrangian as
\begin{equation}
\frac{\alpha^{ut}_{uG \phi}}{\Lambda^2}    \, \mathcal{O}^{ut}_{uG \phi} \quad .
\end{equation}
Considering $\Lambda = 1$ $TeV$ then $\alpha_1$ is defined as
\begin{equation}
\alpha_1 = \frac{\alpha^{ut}_{uG \phi}}{\Lambda^2}    \, TeV^2
\end{equation}
which renders $\alpha_1$ dimensionless. Similar definitions hold for the 
remaining $\alpha_i$ constants. Table~\ref{tab:alphas} shows the relations 
between all the constants shown in the plots and the independent FCNC operators 
in the strong and electroweak sectors.

\begin{table}
\begin{center}
\begin{tabular}{| c c c |}
\hline\hline
Constant & & Operator \\ \hline
$\alpha_1$  & & $\mathcal{O}^{ut}_{uG \phi}$ \\ \hline
$\alpha_2$ & &  $\mathcal{O}^{tu}_{uG \phi}$ \\ \hline
$\alpha_3$ & & $\mathcal{O}^{ut}_{uW \phi}$ \\ \hline
$\alpha_4$ & & $\mathcal{O}^{tu}_{uW \phi}$ \\ \hline
$\alpha_5$ & & $\mathcal{O}^{ut}_{uB \phi}$ \\ \hline
$\alpha_6$ & & $\mathcal{O}^{tu}_{uB \phi}$ \\ \hline
$\alpha_7$&&$\mathcal{O}^{ut}_{\phi u}+\mathcal{O}^{tu}_{\phi u}$\\\hline
$\alpha_8$&&$\mathcal{O}^{(3,tu)}_{\phi q}+\mathcal{O}^{(3,ut)}_{\phi q}$ 
\\ \hline
\end{tabular}
\caption{Relation between the constants presented in the plots and the 
independent FCNC operators in the strong and electroweak sectors. 
\label{tab:alphas}}
\end{center}
\end{table}

\begin{figure}[!h]
\epsfig{file=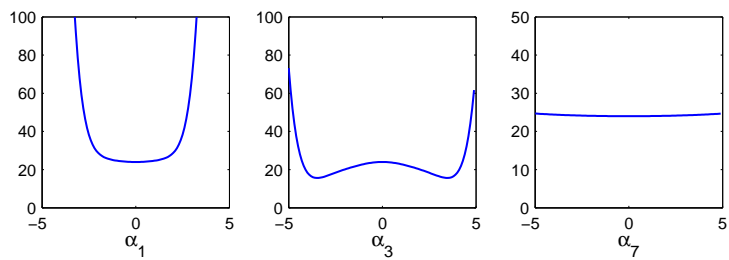, width=10.00cm}
\caption{\label{fig:SE1} The $\chi^2$ as a function of $\alpha_1$, $\alpha_3$ 
and $\alpha_7$ with each parameter taken non-zero at a time. The most 
favoured values are $\alpha_1 = 0$ and $|\alpha_3| \neq 0 $.  There are no 
preferred values for $\alpha_7$. }
\end{figure}

\noindent
We first present our results for the Strong and Electroweak FCNC operators 
(SEFCNC). In Fig.~\ref{fig:SE1} we present the $\chi^2$ as a function of 
$\alpha_1$, $\alpha_3$ and $\alpha_7$, keeping only one of the coefficients 
non-zero at a time. These three curves are representative of the  
$\chi^2$ distribution behaviour for the complete set of SEFCNC operators. In 
fact, we can group operators $\alpha_1$ and $\alpha_2$ as for both $\alpha_1=0$
and  $\alpha_2=0$ are the most favoured values.  The operators that are 
preferentially non zero when taken one at a time are $\alpha_3 $, $\alpha_4$,
$\alpha_5$ and $\alpha_6$. In this case the preferred values are close to 
$\alpha_i = \pm 4$ (see $\alpha_3$ in Fig~\ref{fig:SE1}). Finally both 
$\alpha_7$ and $\alpha_8$ seem to be completely unconstrained as they have 
an almost flat $\chi^2$ distribution for the entire $\alpha_i$ range presented.  
%
\begin{figure}[!h]
\epsfig{file=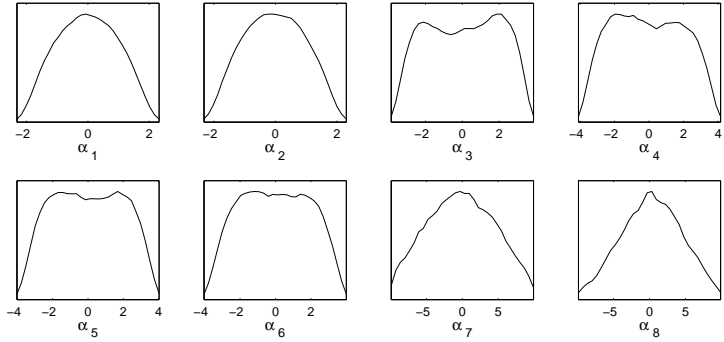, width=10.00cm}
\caption{\label{fig:SE2}One dimensional likelihood distribution of the parameters $\alpha_1$ to $\alpha_8$ after the fit. } 
\end{figure}
\noindent
We have then proceeded to scan over the 8 parameters ($\alpha_i$, $i=1-8$) using 
the MCMC method with flat prior as described in the previous section. The range
for all parameters was chosen to be $-10 < \alpha_i < 10$. The complete set of 
14 experimental observables, presented in the introduction, is used to 
calculate the $\chi^2$ and hence the likelihood. After the likelihood mapping for 
the model, we have obtained the one dimensional likelihood distribution of the 
parameters which is presented in Fig~\ref{fig:SE2}. It is clear from the figure 
that both $\alpha_1$ and $\alpha_2$, the FCNC operators stemming from the 
strong sector, are strongly constrained to be in the range $-2$ to $2$. 
Operators $\alpha_3$ to $\alpha_6$, the chirality-flipping FCNC operators 
coming from the electroweak sector,  have to be in the range $-4$ to $4$. 
Finally the chirality-conserving operators from the electroweak sector, 
$\alpha_7$ and $\alpha_8$ are very mildly constrained and, as we will show 
later, the center-peaked shape of the distribution is only a reflection of the 
correlations of these parameters with the constrained ones.
\begin{figure}[!h]
\epsfig{file=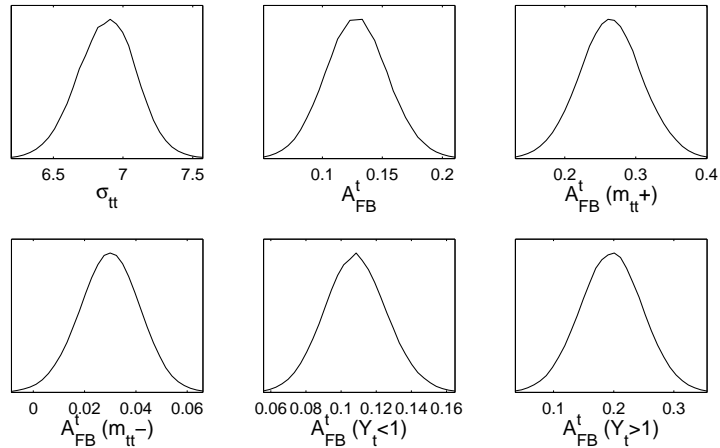, width=10.00cm}
\caption{\label{fig:SE3} One dimensional likelihood distribution of the total cross section and all asymmetries after the fit. } 
\end{figure}
In order to understand how well the model fits into the experimental 
observables we present in fig~\ref{fig:SE3} the one dimensional likelihood 
distribution of the observables after the fit. Considering the central values 
of the observables presented in the introduction, we conclude that the SEFCNC 
set of operators prefer a lower value of the cross section while generating 
suitable values for the total forward-backward asymmetry.  The values for 
$A_{FB}^t(m_{tt}^+)$ and $A_{FB}^t(y_t^+)$ are below their experimental central 
values but well within the error bands. This shows that there is some 
compromise between the values of the parameters in the attempt to fit all 
observable simultaneously giving rise to a slight difference between the input 
observables and the ones originated from the posterior probability distribution
functions (pdfs).

\begin{figure}[!h]
\epsfig{file=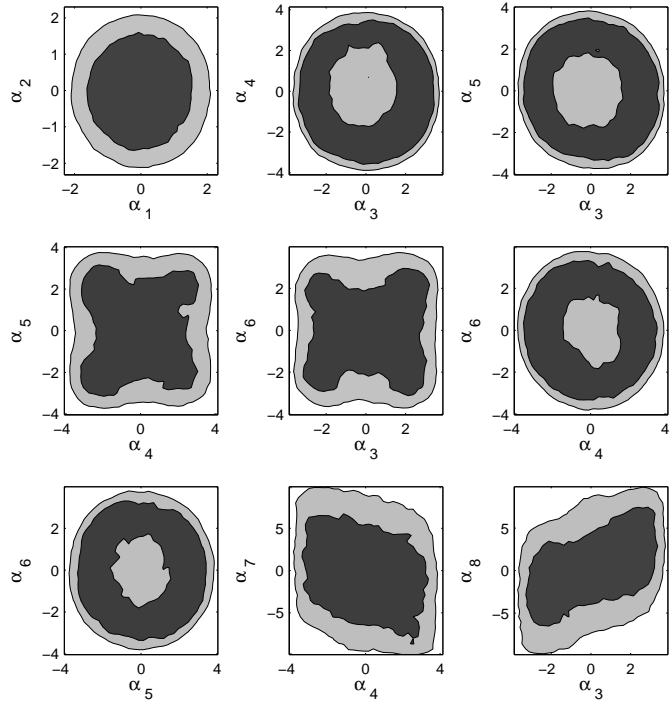, width=9.00cm}
\caption{\label{fig:SE4} Two-dimensional correlation plots for the strong and 
electroweak FCNC parameters. Only the most relevant correlations are shown. The 
shaded areas are the ones where the values of $\alpha_i$ reach their highest 
probability - the darker ones represent 95 \% CL while the lighter ones are 
for 68 \% CL } 
\end{figure}
\noindent
We now move to the study of the possible two-dimensional correlation between 
pairs of parameters. In Fig.~\ref{fig:SE4} we present the two-dimensional 
correlation plots for the most representative scenarios. It is clear from the 
figure that there is no correlation between $\alpha_1$ and $\alpha_2$. 
Furthermore, these operators are very strongly constrained. On the other hand, 
there are several pairs of values that cannot be zero simultaneously. This is 
the case of ($\alpha_3$,$\alpha_4$) -- the ones from SU(2), 
($\alpha_5$,$\alpha_6$) -- the ones from U(1) and ($\alpha_3$,$\alpha_5$), 
($\alpha_4$,$\alpha_6$)  -- these are the U(1) and SU(2) combination where the 
indices of the operators $O^{ij}$ are the same as for example in 
$\mathcal{O}^{ut}_{uW \phi}$ and  $\mathcal{O}^{ut}_{uB \phi}$. For the pairs  
($\alpha_4$,$\alpha_5$) and ($\alpha_3$,$\alpha_6$) the preferred values lie 
in the region $\alpha_4^2 = \alpha_5^2$ and $\alpha_3^2 = \alpha_6^2$ 
respectively. This happens to the combination of SU(2) and U(1) operators with 
the $ij$ indices exchanged. Finally operators $\alpha_7$ and $\alpha_8$ do not 
appear to be much constrained when plotted against the remaining operators. There 
are however mild correlations - if we take for instance the pair 
($\alpha_4$,$\alpha_7$) it is clear that for $\alpha_4 <0$, $\alpha_7$ prefers 
to be positive and if $\alpha_4 >0$, $\alpha_7$ prefers to be negative. 

\begin{figure}[!h]
\epsfig{file=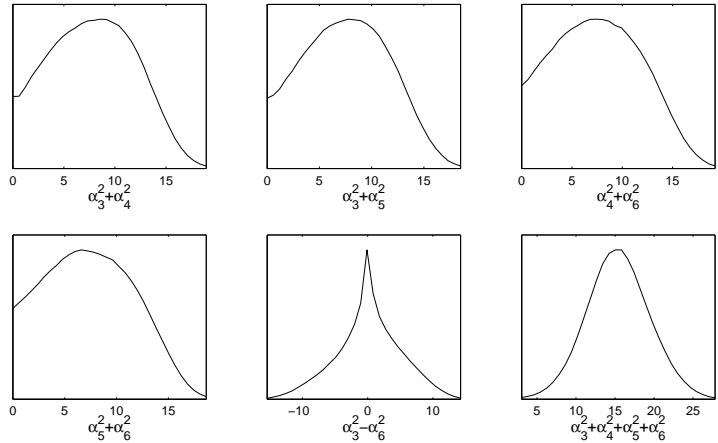, width=10.00cm}
\caption{\label{fig:SE5}Likelihood distributions for the most relevant combination of parameters. }
\end{figure}

\noindent
With the hints from Fig.~\ref{fig:SE4} about which parameters prefer to be 
non-zero after the fit, we have tried to understand  if one could make a more 
strong statement about the appearance of new physics related to the Strong and 
Electroweak dimension six FCNC operators. We note that the contributions of 
$\alpha_7$ and $\alpha_8$ are irrelevant because the change in likelihood is 
very small when these parameters are varied as shown in figures \ref{fig:SE1} and \ref{fig:SE2}. On the other
hand, $\alpha_1$ and $\alpha_2$ can lead to a large change in the likelihood - the 
preferred points are therefore $\alpha_1=0=\alpha_2$. Hence, we look at the most relevant 
combinations of the remaining parameters. The likelihood distributions for 
those combinations are shown in Fig.~\ref{fig:SE5}.  It is clear that all 
the correlated pairs of parameters prefer to be non-zero 
simultaneously, like for instance ($\alpha_3$,$\alpha_4$), which have a peak between 
5 and 10. Again, the likelihood plot  for $\alpha_3^2 - \alpha_6^2$ peaks
at $0$ indicating that $\alpha_3^2=\alpha_6^2$ is the preferred parameter
choice as also seen in Fig.~\ref{fig:SE4}. 
However, the most interesting case is the likelihood for 
$\alpha_3^2 + \alpha_4^2 + \alpha_5^2 + \alpha_6^2$ -- in this case we are 
certain that at least one of the four parameters has to be non-zero in order to 
fit the data. This is a very strong statement because it means that new physics 
coming from these operators can help curing the asymmetry discrepancy and in 
order to solve it at least one of the operators has to be present.

\begin{figure}[!h]
\epsfig{file=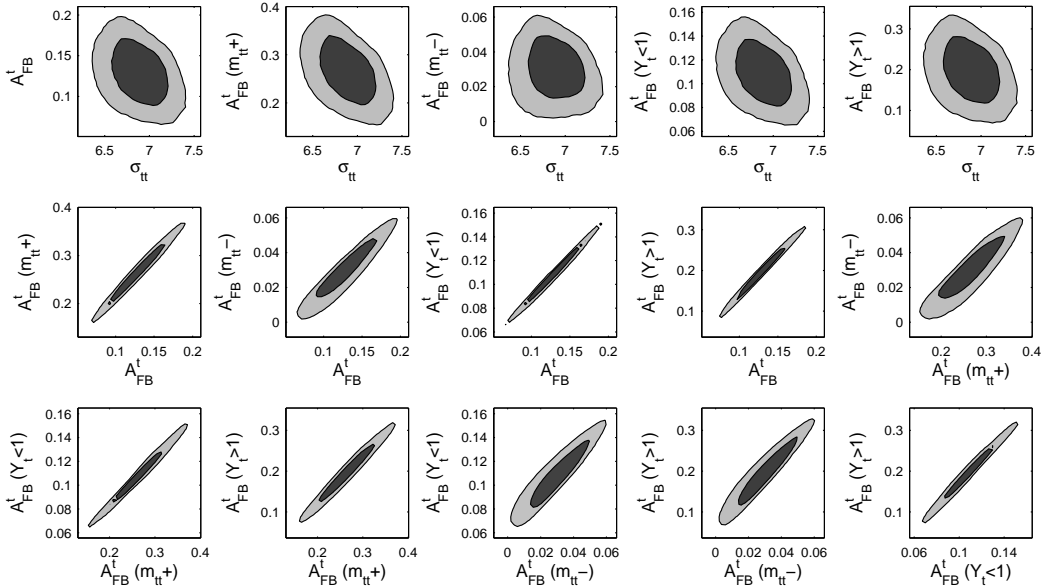, width=14.00cm}
\caption{\label{fig:SE6} Two-dimensional correlations between cross sections 
and asymmetries and between the different asymmetries.}
\end{figure}

In Fig.~\ref{fig:SE6}  we present the two-dimensional correlation between 
several observables after the fit. In the first row one can see that there is a 
negative correlation between asymmetries and total cross section. Hence, to get 
the right asymmetries the cross section moves to its lower preferred value. On 
the other hand, all asymmetries have positive correlations and are highly 
correlated -- if one of them increases the other increases as well.  Therefore, 
there is a tension between cross sections and asymmetries that reflects the 
difficulty of fitting all the observables with the set of SEFCNC operators. 
Nevertheless, a non zero contribution from the operators $\alpha_3$ to 
$\alpha_6$ provides a better fit than the SM one. 

\begin{table}
\begin{center}
\begin{tabular}{||l|r|rr|rr||}\hline\hline
& & \multicolumn{2}{|c|}{68\% interval} & \multicolumn{2}{|c|}{95\% interval}\\
Quantities & Best fit & lower & upper & lower & upper \\ \hline
 $\alpha_{ 1}                                $ & $-0.548 $ & $-1.081 $ & $ 1.066 $ & $-1.797 $ & $ 1.793 $\\  
 $\alpha_{ 2}                                $ & $-0.449 $ & $-1.102 $ & $ 1.053 $ & $-1.812 $ & $ 1.781 $\\  
 $\alpha_{ 3}                                $ & $-2.024 $ & $-2.222 $ & $ 2.293 $ & $-3.257 $ & $ 3.288 $\\  
 $\alpha_{ 4}                                $ & $ 2.913$  & $-2.327 $ & $ 2.250 $ & $-3.443 $ & $ 3.446 $\\  
 $\alpha_{ 5}                                $ & $ 2.403 $ & $-2.210 $ & $ 2.254 $ & $-3.334 $ & $ 3.379 $\\  
 $\alpha_{ 6}                                $ & $ 0.742 $ & $-2.172 $ & $ 2.163 $ & $-3.330 $ & $ 3.317 $\\  
 $\alpha_{ 7}                                $ & $-3.318 $ & $-5.257 $ & $ 4.901 $ & $-8.763 $ & $ 8.536 $\\  
 $\alpha_{ 8}                                $ & $-3.146 $ & $-4.106 $ & $ 4.647 $ & $-7.931 $ & $ 8.148 $\\  
 \hline
 $\sigma_{tt}                                $ & $ 6.817 $ & $ 6.670 $ & $ 7.093 $ & $ 6.453 $ & $ 7.299 $\\  
 $A^{t\bar t}_{FB}                            $ & $ 0.153 $ & $ 0.102 $ & $ 0.155 $ & $ 0.078 $ & $ 0.181 $\\  
 $A^t_{FB}(m_{t\bar t}-)$  & $ 0.044 $ & $ 0.018$ & $ 0.041 $ & $ 0.006 $ & $ 0.053 $\\  
 $A^t_{FB}(m_{t\bar t}+)                        $ & $ 0.310 $ & $ 0.220 $ & $ 0.309 $ & $ 0.177 $ & $ 0.354 $\\  
 $A_{FB}^t (Y_{t}<1)                         $ & $ 0.126 $ & $ 0.090 $ & $ 0.126 $ & $ 0.074 $ & $ 0.144 $\\  
 $A_{FB}^t (Y_{t}>1)                         $ & $ 0.245 $ & $ 0.143 $ & $ 0.248 $ & $ 0.093 $ & $ 0.299 $\\  
 \hline
 $\alpha_3^2+\alpha_4^2+\alpha_5^2+\alpha_6^2$ & $ 18.91 $ & $ 11.43 $ & $ 19.39 $ & $ 7.50 $ & $ 23.46 $\\  
\hline
\end{tabular}
\end{center}
\caption{\label{tab:fcnc}Best fit values and the Bayesian confidence intervals
(BCI) for parameters and the observables.}
\end{table}
In Table.~\ref{tab:fcnc}, we show the best fit values along with $68\%$ and
$95\%$ Bayesian confidence intervals (BCI) for all the parameters and selected
observables. The BCIs are derived from the one-dimensional marginalized 
distributions, as shown  in figures \ref{fig:SE1} and \ref{fig:SE2}, while the best fit point is the one
with least $\chi^2=14.2$. Thus, the best fit point does not need to be at the center 
of the marginalized BCIs. For the SM we have $\chi^2=24.0$ and it is the 
sizeable contributions from $\alpha_3,...,\alpha_6$ operators that lead to the
reduction in the $\chi^2$ for our fits. We again note that the combination
$\alpha_3^2+\alpha_4^2+\alpha_5^2+\alpha_6^2 > 7.5$ with $97.5\%$ CL, i.e. it
is almost certainly non-zero. 

We have also listed the posterior BCI for the cross section and the 
asymmetries in table~\ref{tab:fcnc}. The best fit value of the total 
cross section, and also the $95\%$ BCI, are somewhat smaller than the measured
central value. The same trend is observed for all the asymmetries except for the integrated 
asymmetry $A^{t\bar t}_{FB}$ which is correctly reproduced and the $A_{FB}^t(m_{tt}-)$
asymmetry which is most likely positive in our model. As previously discussed, the
reduction of the cross section values and asymmetries is a result of the negative correlations 
between them.

\subsection{Four fermion operators}

\begin{figure}[!h]
\epsfig{file=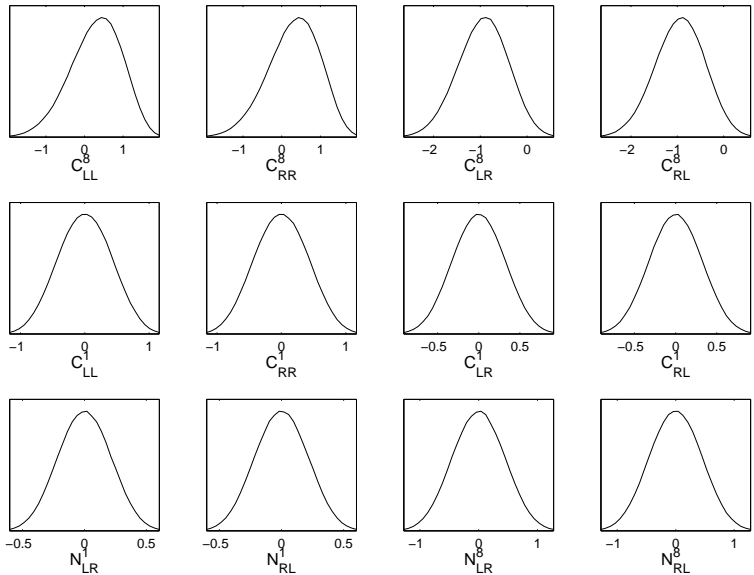, width=10.00cm}
\caption{\label{fig:4F1} Likelihood distribution for the parameters of the four-fermion Lagrangian, after the fit.} 
\end{figure}

\noindent
We now turn our attention to the four-fermion (4F) effective Lagrangian. We 
should start by mentioning that recently~\cite{Shao:2011wa}, a complete 
calculation of the forward-backward asymmetry and of the total cross section 
of top quark pair production induced by 4F-operators was performed for the 
Tevatron up to $O(\alpha^2_s/\Lambda^2)$. The results show that next-to-leading 
order QCD corrections can change both the asymmetry and the total cross section 
by about 10\%. As discussed in section~\ref{sec:4F}, there are a total of 12 
independent operators for the study of $t \bar{t}$ production and under the 
conditions described previously which mainly means we are only considering the 
u-quark contribution in the initial state.  We have scanned linearly over the 
12  parameters from the 4F-Lagrangian using the MCMC method. The range chosen 
for all parameters was again from $-10$ to $10$. In Fig.~\ref{fig:4F1} we 
present the likelihood distribution for all the 4F parameters, after the fit. 
A few comments are in order. First, operators in one row can only interfere 
with parameters in the same row.  Second, only parameters in the first row 
interfere with the SM Lagrangian and consequently the main contribution for the 
asymmetry has to come from the parameters presented in the first row. This is 
clear from the plot as the four distributions in the first row are the only 
asymmetric ones - all other parameters in the following two rows have not only 
symmetric distributions but they show that the preferred value of these 
parameters is zero.  
\begin{figure}[!h]
\epsfig{file=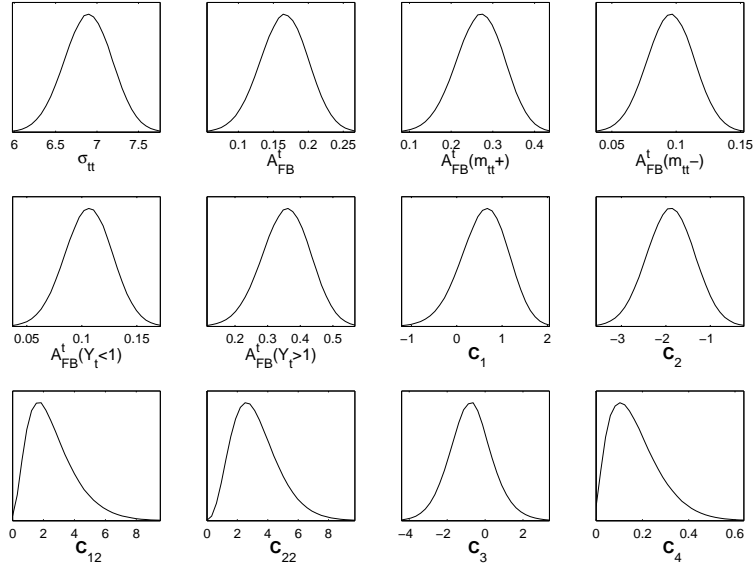, width=10.00cm}
\caption{\label{fig:4F2}Likelihood distribution for the total cross section and for the asymmerties and for the six independent combinations of the 4F Lagrangian parameters, after the fit. } 
\end{figure}
However, in the case of 4F operators the cross sections and the asymmetries 
depend only on six combinations of the parameters. Therefore we have decided to 
present in Fig.~\ref{fig:4F2} the likelihood distribution for those 
combinations together with the likelihood distributions of the total cross 
section and a few selected asymmetries. The relation between the new parameters 
and the original ones present in the 4F Lagrangian is
\begin{eqnarray}
& & C_{1 }  = C^8_{LL} + C^8_{RR} \nonumber \\ 
& & C_{2 }  = C^8_{LR} + C^8_{RL} \nonumber \\ 
& & C_{12}  = (C^8_{LL})^2 + (C^8_{RR})^2  + \frac{9}{2} \left[ (C^1_{LL})^2 + (C^1_{RR})^2 \right] \nonumber \\ 
& & C_{22}  = (C^8_{LR})^2 + (C^8_{RL})^2  + \frac{9}{2} \left[ (C^1_{LR})^2 + (C^1_{RL})^2 \right] \nonumber \\ 
& & C_{3 }  = C^8_{LL}C^8_{LR} + C^8_{RR}C^8_{RL} + \frac{9}{2} \left[ C^1_{LL}C^1_{LR} + C^1_{RR}C^1_{RL} \right] \nonumber \\ 
& & C_{4 }  =  (N^1_{LR})^2 + (N^1_{RL})^2  + \frac{2}{9} \left[ (N^8_{LR})^2 + (N^8_{RL})^2 \right] \quad .
\end{eqnarray}
It is clear that the experimental observables are well described by the fit. 
Regarding the parameters, the most relevant fact, that could already be 
inferred form Fig.~\ref{fig:4F2}, is that $C_2$ prefers to be non-zero and, 
for the same reason, the likelihood of both $C_{12}$ and $C_{22}$ peaks at 1.  
\begin{figure}[!h]
\epsfig{file=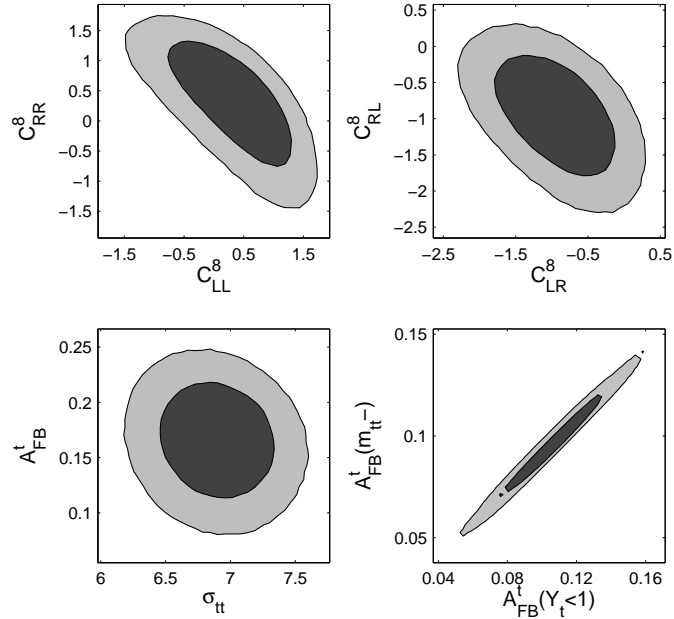, width=9.00cm}
\caption{\label{fig:4F3}Two-dimensional correlations between the parameters that can give a significant contribution to the asymmetry. Also shown are typical examples of the correlations between cross section and asymmetries and between two asymmetry observables.}
\end{figure}
A similar trend can now be seen in the two-dimensional correlations presented 
in Fig.~\ref{fig:4F3}. It is clear that at 95 \% CL the value zero is 
excluded in the top right plot. In the top left plot the value zero is still 
inside the 95 \% CL contour. Regarding the correlations between cross 
section and asymmetries, and between pair of asymmetries, after the fit, the 
general trend is very similar to the one presented in the previous section for
the strong and electroweak FCNC operators. Therefore we will make no further 
comments on those correlations.

\begin{figure}[!h]
\epsfig{file=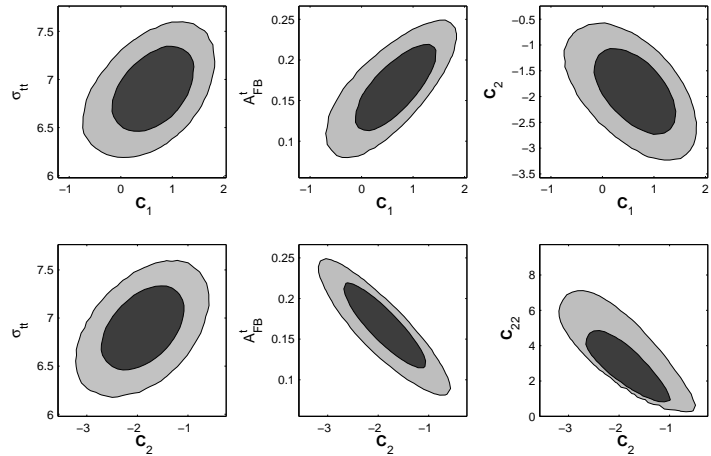, width=10.00cm}
\caption{\label{fig:4F4}Two-dimensional correlations between the parameters $C_1$ and $C_2$ and the total cross section, total asymmetry and the parameter $C_{22}$.} 
\end{figure}

\noindent
In Fig.~\ref{fig:4F4} we present two dimensional correlations between $C_1$ 
and $C_2$ and the total cross section, total asymmetry and the parameter 
$C_{22}$. We see that while $C_1$ is positively correlated with both the cross 
section and the asymmetry,  $C_2$ is positively correlated with the cross 
section but negatively correlated with the asymmetry. Furthermore $C_1$, $C_2$ 
and $C_{22}$ are all negatively correlated with each other. Finally, we conclude 
that either $C_2$, $C_{22}$ or both have to be non-zero which is not surprising 
given the relations presented in eqs. (3.5).

\begin{table}
\begin{center}
\begin{tabular}{||l|r|rr|rr||}\hline\hline
& & \multicolumn{2}{|c|}{68\% interval} & \multicolumn{2}{|c|}{95\% interval}\\
Quantities & Best fit & lower & upper & lower & upper \\ \hline
$C^8_{LL} $ & $ 0.915$ & $-0.385$ & $ 0.968$ & $-1.119$ & $1.466$\\
$C^8_{RR} $ & $ 0.418$ & $-0.368$ & $ 0.980$ & $-1.104$ & $1.476$\\
$C^8_{LR} $ & $-0.934$ & $-1.487$ & $-0.406$ & $-2.031$ & $0.064$\\
$C^8_{RL} $ & $-0.963$ & $-1.488$ & $-0.406$ & $-2.035$ & $0.071$\\
$C^1_{LL} $ & $-0.136$ & $-0.420$ & $ 0.420$ & $-0.794$ & $0.792$\\
$C^1_{RR} $ & $ 0.002$ & $-0.422$ & $ 0.419$ & $-0.795$ & $0.793$\\
$C^1_{LR} $ & $-0.082$ & $-0.316$ & $ 0.316$ & $-0.606$ & $0.606$\\
$C^1_{RL} $ & $ 0.049$ & $-0.316$ & $ 0.318$ & $-0.606$ & $0.607$\\
$N^1_{LR} $ & $ 0.057$ & $-0.212$ & $ 0.212$ & $-0.405$ & $0.405$\\
$N^1_{RL} $ & $-0.036$ & $-0.212$ & $ 0.212$ & $-0.405$ & $0.404$\\
$N^8_{LR} $ & $ 0.070$ & $-0.442$ & $ 0.441$ & $-0.848$ & $0.846$\\
$N^8_{RL} $ & $ 0.040$ & $-0.446$ & $ 0.443$ & $-0.852$ & $0.850$\\
\hline
$\sigma_{tt} $ & $ 7.054$ & $ 6.601 $ & $ 7.181$ & $ 6.315$ & $7.453$\\
$A^{t\bar t}_{FB}$ & $ 0.191$ & $ 0.131$ & $ 0.199$ & $ 0.096$ & $0.231$\\
$A_{FB}^t(m_{tt}-)$ & $ 0.107$ & $ 0.077$ & $ 0.114$ & $ 0.059$ & $0.132$\\
$A_{FB}^t(m_{tt}+)$ & $ 0.321$ & $ 0.211$ & $ 0.327$ & $ 0.151$ & $0.379$\\
$A_{FB}^t(Y_t<1) $ & $ 0.121$ & $ 0.084$ & $ 0.128$ & $ 0.063$ & $0.148$\\
$A_{FB}^t(Y_t>1) $ & $ 0.420$ & $ 0.281$ & $ 0.430$ & $ 0.205$ & $0.496$\\
\hline
\end{tabular}
\end{center}
\caption{\label{tab:4F} The table of best fit values for the 4F case along with $68\%$ and $95\%$ BCI.}
\end{table}

In Table.~\ref{tab:4F}, we show the best fit point along with  $68\%$ and 
$95\%$ BCIs. The best fit point is the one with least $\chi^2=6.28$.
As seen in figure~\ref{fig:4F3}, only $C^8_{AB}$ operators have relevant
contributions to both the asymmetries and the cross sections. The weak operators,
$C^1_{AB}$ do not interfere with the SM diagrams, contributing therefore more
to the cross sections and much less to the asymmetries. Thus, they are 
strongly constrained through the measured values of the cross sections. The $N^i_{AB}$ operators 
contribute only to the cross sections and consequently are also strongly
constrained and irrelevant as possible new physics contributions.

Again, due to the negative correlations between the cross section and the 
asymmetries, there is a slight tension in the fits. This leads to a mild 
preference for lower values of the total cross section. The asymmetries, on the
other hand, are reasonably well reproduced. We note that, $A^t_{FB}(m_{tt}-)$
prefers to be positive with 4F operators.

\section{Bounds on the effective operators}

In this section we discuss all possible bounds
on the dimension six effective operators described
in the previous sections. Our goal is to ascertain whether the
values of the couplings multiplying each operator, that
could explain the measured asymmetry discrepancy, are still allowed
by the available experimental data.
We start by considering the dimension
six FCNC operators. In section III B we have concluded
that only effective operators stemming 
from the electroweak sector were likely to fit the Tevatron
data on the top quark better than the SM. 
In fact, it is the sizeable contributions from the 
$\alpha_3,...,\alpha_6$ operators that lead to the
reduction in the $\chi^2$ for our fits. We again note that the combination
$\alpha_3^2+\alpha_4^2+\alpha_5^2+\alpha_6^2 > 7.5$ with $97.5\%$ CL, it
is almost certainly non-zero. Therefore, we have now to focus on the bounds
for operators $\alpha_3$ to $\alpha_6$ to understand if such a high
value of the constants is not in contradiction with experimental data
from other sources.

A very complete analysis on the electroweak FCNC operators was performed
in~\cite{Fox:2007in} using not only all available data from B physics but
also the data from direct FCNC top decays (the later will be updated
in this work) \footnote{Other analysis based on B physics observables and 
electroweak precision constraints were also performed
in~\cite{Drobnak:2011wj} leading to similar conclusions.}.  
The bounds obtained on the operators taken one at
a time are~\cite{Fox:2007in} $\alpha_3^2 < 0.81$,
$\alpha_4^2 < 0.011$ and $\alpha_6^2 < 0.096$ while 
the best bound on $\alpha_5$ was shown to come from
the direct constraint on $BR (t \to q Z)$ and $BR (t \to q \gamma)$. 
Therefore, to satisfy $\alpha_3^2+\alpha_4^2+\alpha_5^2+\alpha_6^2 > 7.5$
one  needs $\alpha_5^5 \approx 6.5$. However, such a value of $\alpha_5$
would imply that $BR (t \to q Z) \approx 3.7 \%$ and $BR (t \to q \gamma) \approx 6.3 \%$.
The most recent direct bounds on $BR (t \to q \gamma) $ and $BR (t \to q Z)$
are the ones from the Tevatron, $3.2 \%$~\cite{Abe:1997fz}  and
from the LHC, $1.1 \%$~\cite{ATLASFCNC}, respectively. Hence, it is clear
that such high values of $\alpha_5$ are disallowed by Tevatron
and LHC data on the direct searches for FCNC top decays with
a photon or a Z-boson in the final state. Furthermore, indirect 
bounds from HERA, where bounds on cross sections are
converted on bounds on the branching ratio, set a limit  
$BR (t \to q \gamma) \lesssim 0.5 \%$~\cite{Abramowicz:2011tv}.
Also, a combine study on B physics and Tevatron data on top 
quark production cross section places an indirect bound on the sum of the 
FCNC branching ratios forcing them to be below the percent level~\cite{Ferreira:2009bf}.
In conclusion, experimental data from very different sources
constrain the operators that could explain the asymmetry
in such a way that we consider that it is very unlikely that
the observed discrepancy could be explained by these operators.

Contrary to the the dimension six FCNC operators, there are no useful
bounds on the four fermion operators involving two top quarks and
this is even more so if the top is right-handed. Therefore, only
the LHC could place constraints on these operators. However the
values of the constants $C_i$ and $C_{ij}$ that could help explain
the discrepancy give an extra cross section that is always below
10 pb even for $\sqrt{s} = 14$ TeV. Hence, given the
error of $t \bar t$ production cross section it is very unlikely
that these operators will be constrained in the near future. 

\section{Discussion and conclusions}

\noindent
In this work we have used a dimension six Lagrangian with FCNC interaction 
together with four-fermion operators to gain some insight in understanding the discrepancy between the 
experimental values obtained for the top pair production asymmetry and the 
corresponding SM predictions. We have build a minimal set of 
operators and we have used an MCMC approach to find the best simultaneous fit 
of all independent operators to the available data.
Our conclusions regarding which operators give the best fit are as follows
\begin{itemize}
\item Strong FCNC operators with coefficients $\alpha_{1,2}$ are most likely close to zero;
\item regarding Electroweak FCNC operators with coefficients $\alpha_3$ to $\alpha_6$ we conclude that one of them must be non-zero;
\item Electroweak FCNC operators with coefficients $\alpha_{7,8}$ are not relevant;
\item bounds on electroweak FCNC operators reveal that it is very unlikely that the asymmetry can come from new physics described by these operators;
\item Four-fermion operators with coefficients $C^8_{AB}$ contribute to the  asymmetries as the ones with coefficients $C^1_{AB}$ give small contributions; 
the 4F combinations $C_1, C_2, C_{12}, C_{22}$ contribute to the asymmetries;
\item Four-fermion operators with coefficients $N^i_{AB}$ contribute to the cross sections only;
\item there is in all cases some tension between cross section and asymmetries when a simultaneous fit to all observables is performed;
\item the contribution of 4F operators to the cross-section at LHC7 is of the order $\pm 1.5$ pb, which is allowed by the present estimates of the cross-section~\cite{Kim:2012gy}.
\end{itemize}

It is important to ask how do we figure out which operators are actually responsible for the asymmetry. To that end we note that the asymmetry, although called forward-backward, is actually a $C$-odd and for $CP$ conserving interaction that can therefore be created by $P$-odd interactions as well.
Further, our operators contribute to the asymmetry in two ways: kinematically and dynamically. The $t$-channel diagrams with FCNC interaction naturally originate more top quarks in the direction of the incoming $u$-quark leading to a positive asymmetry as measured by the CDF collaboration. This coupling does not need to be chiral to produce the required asymmetry, although our operators are chiral. For the 4F case, there is no such kinematical asymmetry generation and it is dominantly generated by the unbalance between left and right chiral operators interfering with the SM contribution. Thus, in both cases, we have the presence of chiral interaction, which also incarnates in the form of polarization of the produced top-quark. Hence, a study of such polarization effects~\cite{Choudhury:2010cd} as a function of rapidity will be able to provide a probe of possible new physics. Further, our operators are also constrained by the B-physics observables and for simplicity we have not accounted for them in our MCMC. We have nevertheless used the constraints from B-physics to conclude that FCNC operators are unlikely to account for the measured asymmetry at the Tevatron 

To conclude, we remark that the $A^{t \bar t}_{FB}$observed at CDF can be casted in terms of dimension-six operators and we need more observables, from top-polarization and B-physics, to constrain them due to the multitude of these operators.


\acknowledgments
We thank Rohini Godbole and Saurabh Rindani for useful discussions in the 
beginning of the collaboration. We also thank the WHEPP-XI organizers for their kind hospitality and for
providing a great atmosphere that was the seed of the work presented in this paper.
RS is supported in part by the Portuguese
\textit{Funda\c{c}\~{a}o para a Ci\^{e}ncia e a Tecnologia} (FCT)
under contract PTDC/FIS/117951/2010, by an
FP7 Reintegration Grant, number PERG08-GA-2010-277025
and by PEst-OE/FIS/UI0618/2011. MW is supported by
FCT under contract SFRH / BD /45041/2008.

\end{document}